\documentclass[11pt]{article}
\usepackage{fullpage}
\usepackage{lscape}
\usepackage[dvips]{epsfig}

\def\##1{{\bf #1}}
\def\=#1{{\cal #1}}

\def\eps{\epsilon}
\def\epso{\epsilon_0}
\def\muo{\mu_0}
\def\ko{k_0}

\def\.{\mbox{ \tiny{$^\bullet$} }}

\def\ux{\#{u}_x}
\def\uy{\#{u}_y}
\def\uz{\#{u}_z}

\def\le{\left(}
\def\ri{\right)}
\def\les{\left[}
\def\ris{\right]}
\def\lec{\left\{}
\def\ric{\right\}}

\def\c#1{\cite{#1}}

\def\r#1{(\ref{#1})}

\def\epsa{\eps_a}
\def\epsb{\eps_b}
\def\mua{\mu_a}
\def\mub{\mu_b}

\def\fz{\les\#f(z)\ris}
\def\Pz{\=P(z)}

\begin{document}

\noindent {\Large {\emph {\bf Restricted equivalence of paired epsilon--negative
and mu--negative layers to a negative phase--velocity material ({\em alias} left--handed
material) }}}

\bigskip

\noindent {\bf Akhlesh Lakhtakia}\footnote{Corresponding
author. Tel: +1 814 863 4319; Fax: +1 814 865 9974; E--mail:
AXL4@PSU.EDU} \bigskip

\noindent  CATMAS~---~Computational and Theoretical Materials Science
Group\\ Department of Engineering Science and Mechanics\\
Pennsylvania State University, University Park, PA 16802--6812, USA
\bigskip

\noindent {\bf Clifford M. Krowne}\bigskip

\noindent Microwave Technology Branch \\
Electronic Science \& Technology Division\\
Naval Research Laboratory, Washington, DC 20375--5347, USA

\bigskip

\noindent {\bf Abstract:}
The time--harmonic electromagnetic responses of (a) a bilayer
made of an epsilon--negative layer and a mu--negative layer,
and (b) a single layer of a negative phase--velocity material
are compared. Provided all layers are electrically thin, a restricted
equivalence between (a) and (b) exists. The restricted equivalence
depends on the linear polarization state and the transverse wavenumber.
Implications for perfect lenses and parallel--plate waveguides are considered.


\section{Introduction}

This communication is inspired by the ongoing 
spate of papers
published on the inappropriately designated {\em left-handed 
materials\/} which are macroscopically
homogeneous and display negative
phase velocities in relation to the time--averaged
Poynting vector, but are not chiral \cite{LMW}.
Nominally, such a material is deemed
to possess a relative permittivity scalar
$\eps_r =\eps_r' + i \eps_r''$ and a relative permeability scalar 
$\mu_r=\mu_r'+i\mu_r''$, both dependent
on the angular frequency $\omega$, such that both
 $\eps_r'<0$ and $\mu_r'<0$
 in some spectral
regime.\footnote{The
condition for the phase velocity and the time--averaged Poynting vector to be
oppositely directed is $\le\vert\eps_r\vert-\eps_r'\ri\le\vert\mu_r\vert-
\mu_r'\ri > \eps_r''\mu_r''$, which
 permits~---~more generally~---~$\eps_r'$
and/or  $\mu_r'$ to be negative \c{MLW}. An $\exp(-i\omega t)$
time--dependence having been assumed here, $\eps_r'' >0$
and $\mu_r'' >0$ at all $\omega>0$ for all passive
materials.}
Originally conceived by Veselago \c{Ves}, these materials
have been artificially realized quite recently \c{SSS,PGLKT}.
Their fascinating electromagnetic properties can have technological
implications of massive proportions \c{Pen}, but those implications
remain speculative at this time. Although these materials have been
variously named \c{LMW}, we prefer the name {\em negative phase--velocity\/} (NPV)
materials as the least ambiguous of all extant names.

Using transmission--line analysis and lumped parameters, Alu and
Engheta \c{Engh} have recently suggested a new route to realizing NPV materials:
Take a thin layer of an {\em epsilon--negative\/}
(EN)  material:  it has a negative real permittivity scalar but a  positive
real permeability scalar.  Stick it to a
thin layer of a {\em mu--negative\/} (MN) material, which
has a negative real permeability scalar and a positive real permittivity scalar.
Provided the two layers are sufficiently thin, the paired EN--MN layers {\em
could\/} function effectively as a NPV material. The clear attraction of
this scheme is
that EN and MN layers are easier to manufacture, very likely,
than the NPV materials fabricated thus far \c{SSS,PGLKT}.

Our objective here is to examine the suggested scheme using continuum field
theory, and to establish a restricted equivalence of an
EN--MN bilayer to a NPV material. Implications for parallel--plate
waveguides \c{Engh} and perfect lenses \c{P2}
are deduced therefrom.

A note about notation: Vectors are in boldface, 
 column vectors are boldface and enclosed in square 
brackets, while
matrixes are denoted by Gothic letters; $\epso$
and  $\muo$
are the free--space permittivity and permeability, respectively; and
$\ko=\omega(\epso\muo)^{1/2}$ is the free--space wavenumber.
  A cartesian 
coordinate system
is used, with $\ux$, $\uy$ and $\uz$ as the
unit vectors.

\section{Bilayer Theory in Brief}
Consider the layers $0< z< d_a$ and $d_a< z< d_a +d_b$.
Their constitutive relations are as follows:
\begin{equation}
\left.\begin{array}{l}
\#D(\#r) = \epso\epsa\,\#E(\#r)\\
\#B(\#r)=\muo\mua\,\#H(\#r)
\end{array}
\ric\,,\qquad 0<z< d_a\,,
\end{equation}
\begin{equation}
\left.\begin{array}{l}
\#D(\#r) = \epso\epsb\,\#E(\#r)\\
\#B(\#r)=\muo\mub\,\#H(\#r)
\end{array}
\ric\,,\qquad d_a<z< d_a+d_b\,.
\end{equation}
The constitutive parameters present in the foregoing equations
are complex--valued with positive imaginary parts (as befits any
passive medium). The two half--spaces on either sides
of the bilayer are vacuous.

Without loss of generality, the electromagnetic  field phasors everywhere can be 
written as \c{K1}
\begin{equation}
\left.\begin{array}{l}
\#E(\#r) = \tilde{\#e}(z)\, e^{ i\kappa x}\\
\#H(\#r) = \tilde{\#h}(z) \,e^{ i\kappa x}
\end{array}\ric\,, \quad -\infty<z<\infty\,,
\end{equation}
where the transverse wavenumber $\kappa\in \les 0,\,\infty\ri$.
The
fields inside the bilayer must follow the 4$\times$4 matrix
ordinary differential equation \c{K1,L1}
\begin{equation}
\label{diffeq}
\frac{d}{dz}\,\fz = i\Pz\thinspace\fz\,,\quad 0<z<d_a+d_b\,.
\end{equation}
In this equation,
$\fz ={\rm col}\, \les\tilde {e}_x(z),\, \tilde{e}_y(z),\,
\tilde{h}_x(z),\, \tilde{h}_y(z)\ris$
is a column vector, while the 4$\times$4 matrix function
$\Pz$ is piecewise uniform as 
\begin{equation}
\Pz = \lec
\begin{array}{l}
{\=P}_a\,,\qquad 0 < z < d_a\,\\
{\=P}_b\,,\qquad d_a < z < d_a + d_b\,
\end{array}\right.\,,
\end{equation}
where
\begin{equation}
{\=P}_{a,b}=
\les \begin{array}{cccc}
0  &  0  &  0  &-\,\frac{\kappa^2}{\omega\epso\eps_{a,b}}+\omega\muo\mu_{a,b}\\
0 & 0 & -\omega\muo\mu_{a,b} & 0 \\
0 & -\omega\epso\eps_{a,b} +\frac{\kappa^2}{\omega\muo\mu_{a,b}} & 0 & 0\\
\omega\epso\eps_{a,b} & 0 & 0 & 0
\end{array}\ris\,.
\end{equation}
The only nonzero elements of the matrixes $\=P_{a,b}$
appear on their antidiagonals, of which the $(2,3)$ and the $(3,2)$
elements are relevant to $s$--polarized fields, and the $(1,4)$ and the $(4,1)$
elements
to the $p$--polarized fields.

The solution of \r{diffeq} is straightforward, because the
matrix $\Pz$ is piecewise uniform \c{Hoch}. Thus, the algebraic relation
\begin{equation}
\label{sys}
\les\#f(d_b+d_a)\ris =
e^{i{\=P}_bd_b}\,e^{i{\=P}_ad_a}\,\les\#f(0)\ris
\end{equation}
is sufficient to solve both reflection/transmission problems
as well as guided--wave propagation problems. The
two matrix exponentials on the right side
of \r{sys} cannot be interchanged~---~unless
the matrixes $\=P_a$ and $\=P_b$ also commute,
which is possible with dissimilar materials {\em only\/}  in quite
special circumstances \c{L1,L2}.

\section{Analysis}
Matrixes $\=P_{a,b}$ have $\pm \sqrt{\ko^2\eps_{a,b}\mu_{a,b}-\kappa^2}
=\pm\alpha_{a,b}$
as their eigenvalues. Provided that $\vert \alpha_{a,b}\vert\,d_{a,b} \ll 1$
(i.e., both layers are electrically thin),
the approximations
\begin{equation}
e^{i\=P_{a,b}d_{a,b}} \simeq \=I  + i\=P_{a,b}d_{a,b}
\end{equation}
can be made, with $\=I$ as the 4$\times$4 identity matrix. Then
\begin{equation}
\label{ab}
e^{i{\=P}_bd_b}\,e^{i{\=P}_ad_a} \simeq
\=I + i\=P_ad_a + i\=P_bd_b \simeq
e^{i{\=P}_ad_a}\,e^{i{\=P}_bd_b}\,,
\end{equation}
and the two layers in the bilayer
 can be interchanged without significant effect \c{Reese}.

Let us now consider a single layer of relative permittivity $\eps_{eq}$,
relative permeability $\mu_{eq}$ and thickness $d_{eq}$. Quantities
$\=P_{eq}$ and $\alpha_{eq}$ can be defined in analogy to
$\=P_a$ and $\alpha_a$. Two thickness ratios are defined as
\begin{equation}
p_{a,b} = \frac{d_{a,b}}{d_{eq}}  \geq 0\,,
\end{equation}
in order to compare the single layer with the previously described
bilayer. There is no hidden restriction on the non--negative real
numbers $p_a$ and $p_b$.

Provided that $\vert \alpha_{eq}\vert
\,d_{eq} \ll 1$ (i.e., the single layer is
electrically thin as well), the approximation
\begin{equation}
\label{eq}
e^{i\=P_{eq}d_{eq}} \simeq \=I +i\=P_{eq}d_{eq}
\end{equation}
can be made. Equations \r{ab} and \r{eq} permit us to establish
the following equivalences between a bilayer and a single
layer:
\begin{itemize}
\item[(i)] {\em s--polarization:\/} The only nonzero field components
are $E_y$, $H_x$ and $H_z$. Therefore, the equality of the
$(2,3)$ elements of $p_a\=P_a+p_b\=P_b$
and $\=P_{eq}$, and likewise
of the $(3,2)$ elements,  has to be guaranteed for equivalence; thus,
the equations
\begin{eqnarray}
&&
p_a\mu_a+p_b\mu_b = \mu_{eq}\,,
\\[5pt]
&&
p_a\eps_a+p_b\eps_b - \le\frac{\kappa}{\ko}\ri^2
\le \frac{p_a}{\mu_a} +\frac{p_b}{\mu_b}\ri
= \eps_{eq} - \le\frac{\kappa}{\ko}\ri^2
 \frac{1}{\mu_{eq}}\,
 \end{eqnarray}
 have to solved for $\eps_{eq}$ and $\mu_{eq}$. We conclude therefrom
 that,
 for a given value of $\kappa$
and subject to the thickness restrictions
$\vert\alpha_{a,b,eq}\vert\,d_{a,b,eq}\stackrel{<}{\sim}  0.1$,
the bilayer and a single layer are equivalent with respect to the
transformation of the
$x$-- and $y$-- components of the fields from one exterior face
to the other exterior face if
\begin{eqnarray}
\nonumber
\eps_{eq} &=& p_a \eps_a + p_b \eps_b\\
&&-\le \frac{\kappa}{\ko}\ri^2
\les\frac{p_ap_b(\mu_a-\mu_b)^2}{\mu_a\mu_b(p_a\mu_a+p_b\mu_b)}
+\frac{(p_a+p_b+1)(p_a+p_b-1)}{p_a\mu_a+p_b\mu_b}\ris\,,
\\
\mu_{eq} &=& p_a\mu_a+p_b\mu_b\,.
\end{eqnarray}

\item[(ii)] {\em p--polarization:\/} The only nonzero field components
being $H_y$, $E_x$ and $E_z$,  the equality of the
$(1,4)$  elements of $p_a\=P_a+p_b\=P_b$
and $\=P_{eq}$ suffices, along with the equality of
the $(4,1)$ elements of the two matrixes. For a given value of $\kappa$
and subject to the thickness restrictions
$\vert\alpha_{a,b,eq}\vert\,d_{a,b,eq}\stackrel{<}{\sim} 0.1$,
the bilayer and a single layer are  equivalent  if
\begin{eqnarray}
\nonumber
\mu_{eq} &=& p_a \mu_a + p_b \mu_b\\
&&-\le \frac{\kappa}{\ko}\ri^2
\les\frac{p_ap_b(\eps_a-\eps_b)^2}{\eps_a\eps_b(p_a\eps_a+p_b\eps_b)}
+\frac{(p_a+p_b+1)(p_a+p_b-1)}{p_a\eps_a+p_b\eps_b}\ris\,,
\\
\eps_{eq} &=& p_a\eps_a+p_b\eps_b\,.
\end{eqnarray}
\end{itemize}
Clearly, the constitutive
parameters of the equivalent
layer are functions of both   $p_a$ and $p_b$; and we
must point out that the sum of these two ratios need not equal unity.
Furthermore,
except for normal incidence (i.e., $\kappa = 0$), the constitutive
parameters of the equivalent
layer depend on the incident linear polarization state. Finally,
the constitutive parameters of the equivalent layer change
with the transverse wavenumber $\kappa$. 

The foregoing equations can be manipulated to yield negative
values of both $\eps_{eq}'$ and $\mu_{eq}'$ for either 
\begin{itemize}
\item
an EN--MN bilayer
$\lec \eps_a' <0,\,\eps_b'>0,\,\mu_a'>0,\,\mu_b'<0\ric$ or
\item
a MN--EN bilayer
 $\lec \eps_a' >0,\,\eps_b'<0,\,\mu_a'<0,\,\mu_b'>0\ric$.
 \end{itemize}
An EN--MN bilayer
 is equivalent to a NPV layer for both
 polarization states when $\kappa=0$, provided the
 condition
 \begin{equation}
 \label{delta1}
 \frac{\vert\mu_b'\vert}{\mu_a'} > \frac{p_a}{p_b} > \frac{\eps_b'}{\vert\eps_a'\vert}
 \end{equation}
 holds true. The inequality  \r{delta1}  is applicable
 for a MN--EN bilayer also, if the subscripts $a$ and $b$
 are interchanged therein, i.e.,
  \begin{equation}
 \label{delta2}
 \frac{\vert\mu_a'\vert}{\mu_b'} > \frac{p_b}{p_a} > \frac{\eps_a'}{\vert\eps_b'\vert}\,.
 \end{equation}
 A further specialization of $p_a+p_b=1$ leads to the inequality
  \begin{equation}
 \label{delta3}
 \frac{\vert\mu_b'\vert}{\mu_a' +\vert\mu_b'\vert}
 > p_a > \frac{\eps_b'}{\vert\eps_a'\vert+\eps_b'}
 \end{equation}
 for EN--MN bilayers, and
  \begin{equation}
 \label{delta4}
 \frac{\vert\mu_a'\vert}{\mu_b' +\vert\mu_a'\vert}
 > p_b > \frac{\eps_a'}{\vert\eps_b'\vert+\eps_a'}
 \end{equation}
 for MN--EN bilayers.
 
The inequalities \r{delta1}--\r{delta4} should be adequate for both
$s$-- and $p$--polarization states when $\kappa/\ko \ll 1$.
In general, however, a given EN--MN (or  MN--EN)
 bilayer
is equivalent to a different NPV material for a different linear
polarization state and/or transverse wavenumber.
 Thus, the equivalence
between an EN--MN  (or a MN--EN) bilayer and a NPV layer is
{\em restricted.} 

The restricted equivalence has an interesting implication for
perfect lenses \c{P2}. A perfect lens of thickness $d >0$ is defined
by the fulfillment of the condition
$\les\#f(d)\ris =\les\#f(0)\ris$ for all $\omega$ and $\kappa$. 
Because of dispersion and
dissipation, at best,  this condition is fulfilled approximately.
Let us imagine that the condition is fulfilled by some NPV
constitutive parameters for some $\omega$
and all $\vert\kappa\vert\leq \hat{\kappa}$. Then, the implementation
of  the acceptably imperfect lens
as a cascade of thin EN-MN (or MN--EN) bilayers would require that  the successive
bilayers have different constitutive parameters and that the entry as well
as the exit faces be curved, and even those steps may not suffice.
In contrast, $\kappa$ is fixed for 
any single--mode parallel--plate  waveguide, and so
is the range of operating frequencies; and the emulation of
a NPV material  {\em via\/}
EN--MN (or MN--EN) bilayers may not be onerous.

\end{document}